\documentclass[aps,twocolumn,showpacs,superscriptaddress,prb]{revtex4}

\usepackage{exscale}
\usepackage{graphicx}
\usepackage{amsmath}
\usepackage{latexsym}
\usepackage{amsfonts}
\usepackage{amssymb}
\usepackage{color}
\usepackage{amscd}
\usepackage{bbm}
\usepackage{dcolumn}      
\usepackage{bm}           
\usepackage{pstricks,pst-grad,fancybox,graphics}
\usepackage{enumerate}
\usepackage{textcomp}

\newcommand{\ket}[1]{\ensuremath{| #1 \rangle}}

\newcommand{\ketbra}[1]{\ensuremath{\left| #1 \right\rangle \left\langle #1 \right|}}

\newcommand{\eins}{\ensuremath{\mathbbm 1}}
\newcommand{\unity}{\eins}

\newcommand{\mcF}{\ensuremath{\mathcal{F}}}
\newcommand{\mcH}{\ensuremath{\mathcal{H}}}
\newcommand{\mcL}{\ensuremath{\mathcal{L}}}

\newcommand{\mcS}{\ensuremath{\mathcal{S}}}
\newcommand{\mcW}{\ensuremath{\mathcal{W}}}

\newcommand{\BE}{\begin{equation}}
\newcommand{\EE}{\end{equation}}
\newcommand{\be}{\begin{equation}}
\newcommand{\ee}{\end{equation}}
\newcommand{\bea}{\begin{eqnarray}}
\newcommand{\eea}{\end{eqnarray}}
\newcommand{\bean}{\begin{eqnarray*}}
\newcommand{\eean}{\end{eqnarray*}}
\newcommand{\kommentar}[1]{}

\newcommand{\mean}[1]{\ensuremath{\langle #1 \rangle}}

\newcommand{\bc}{\begin{center}}
\newcommand{\ec}{\end{center}}

\renewcommand{\exp}[1]{\ensuremath{\textrm{exp}\hspace{-.06cm}\left[ #1 \right]}}
\newcommand{\tr}[1]{{\rm Tr}\left[ #1 \right]}

\definecolor{red}{rgb}{1,0,0}

\renewcommand{\vr}{\ensuremath{\varrho}}

\begin{document}

\title[Decay of Entanglement in Quantum Dots]{Modeling the Decay of Entanglement for Electron Spin Qubits in Quantum Dots}

\author{F. Bodoky}
\affiliation{Kavli Institute of Nanoscience, Delft University of Technology, Lorentzweg 1, 2628 CJ Delft, The Netherlands}

\author{O. G\"uhne}
\affiliation{Institut f\"ur Quantenoptik und Quanteninformation,
~\"Osterreichische Akademie der Wissenschaften, 
6020 Innsbruck, Austria,}
\affiliation{Institut f\"ur Theoretische Physik,
Universit\"at Innsbruck, Technikerstra{\ss}e 25,
6020 Innsbruck, Austria}

\author{M. Blaauboer}
\affiliation{Kavli Institute of Nanoscience, Delft University of Technology, Lorentzweg 1, 2628 CJ Delft, The Netherlands}

\date{\today}

\begin{abstract}
We investigate the time evolution of entanglement under various models of decoherence: A 
general heuristic model based on local relaxation and dephasing times, 
and two microscopic models describing decoherence 
of electron spin qubits in quantum dots due to the hyperfine interaction 
with the nuclei. For each of the decoherence models, we investigate and compare how long the entanglement 
can be detected. We also introduce filtered witness operators, which extend the available 
detection time, and investigate this detection time for various 
multipartite entangled states. By comparing the time required for detection with the time required 
for generation and manipulation of entanglement, we estimate for a range of different entangled 
states how many qubits can be entangled in a one-dimensional array of electron spin qubits.
\end{abstract}


\maketitle
\section{Introduction}
Entanglement refers to non-classical correlations between two~\cite{epr35,sch35,bell64,werner} or 
more~\cite{horo,horo07} quantum particles, and the creation of multiparticle entangled states 
constitutes a key step towards quantum computation~\cite{niel00}. In this work, we investigate the 
evolution of entangled states under different models of decoherence: a heuristic model with a broad 
range of applications, and two microscopic models specific for electron spin qubits in quantum 
dots~\cite{han07}. We show how entanglement can be detected, and how fast this needs to be done 
before the states become disentangled by decoherence. We also estimate for four different classes 
of multipartite entanglement which class survives the longest, and how many entangled qubits can be 
generated and detected with actual experimental means and currently known decoherence times. 

We consider decoherence of a {\it local} nature, {\it i.e.} the qubits decohere in an uncorrelated 
way, as is the case in solid-state nanosystems such as electron spin qubits in quantum 
dots~\cite{han07}, various superconducting qubits~\cite{wen06} and other solid-state 
implementations~\cite{han08}: In these systems, the decoherence can be characterized 
phenomenologically~\cite{eng02} by a relaxation time $T_{1}$ and a dephasing time $T_{2}$. The 
microscopic origin of the decoherence is still a matter of intensive research: In this paper 
we discuss some microscopic models for electron spin qubits in quantum 
dots and compare them with this heuristic model. 

Our proposed means to analyze entanglement are so-called witness 
operators~\cite{terhal,lew00,guh02,guh03,guh09}: locally decomposable observables with a positive 
expectation value for all separable states, and a negative expectation value for at least one 
entangled state. The advantage of entanglement witnesses over other methods such as {\it e.g.} full 
state tomography is that they require less measurements, and thus less experimental effort to 
detect and prove the existence of entanglement for a given (mixed) state. Witnesses have 
intensively been used in experiments with photons~\cite{bou04,witPhot} and trapped 
ions~\cite{witIon}, but so far only few theoretical proposals exist for using witness operators in 
solid-state nanosystems~\cite{witSS}.\\

This paper is organized as follows: in Sec.~\ref{sDecMod} we briefly summarize the mechanisms 
influencing the time scales $T_{1}$ and $T_{2}$ for electron spin qubits. We describe two 
theoretical models of dephasing for these qubits and compare them to the heuristic master equation 
model. We also show how we calculate the decoherence of a multipartite state of separated qubits using 
the Lindblad formalism. Next (Sec.~\ref{sSit}) we demonstrate our main ideas and methods using the 
simplest case of two qubits, compare the different models of decoherence and introduce our filtered 
witness operator. In Sec.~\ref{s3Qub} we consider both GHZ- and W-states for three qubits. In 
Sec.~\ref{s4Qub} we do the same for four qubits and consider in addition the cluster and Dicke 
entanglement classes. We also discuss the dependence of the decay of entanglement on the initial 
state. Finally, in Sec.~\ref{sNQub}, we discuss the case of $N$ qubits: We show that the 
entanglement of a specific GHZ-state can theoretically be detected for any finite time, and discuss 
the feasibility of generating and detecting many-qubit entangled states of electron spin qubits 
based on decoherence and operation time scales that have been measured in recent experiments on 
single and double quantum dots.
%
%
%
\section{Decoherence model}\label{sDecMod}
Decoherence is caused by uncontrolled interactions between the qubit and the 
environment~\cite{sch05}. This effect is usually characterized by two time scales, the phase 
randomization time $T_{2}$ (``dephasing time'') and the time $T_{1}$ in which the excited state 
$\ket{1}$ relaxes to the ground state $\ket{0}$ by energy exchange with the environment 
(``relaxation time'')~\cite{eng02}. For electron spin qubits (as for most solid-state 
qubits) the dephasing time is much shorter than the relaxation time, $T_{2}\ll T_{1}$, and is 
therefore the dominant time scale for the loss of quantum correlations. In this section, we 
consider both a simple exponential model of decoherence based on these two time scales, as well as use two microscopic 
descriptions of dephasing for electron spin qubits in quantum dots to derive more sophisticated time 
evolutions of decoherence.

We start by briefly discussing decoherence mechanisms for electron spin qubits. 
The original idea by Loss and DiVincenzo~\cite{loss98} proposed to confine single electrons in a 
quantum dot (an island of charge in a two-dimensional electron gas) and apply a magnetic field to 
split the degeneracy of the spin-up and spin-down states, thus creating a two-level system serving 
as carrier for quantum information: an electron spin qubit. Two electron spin qubits interact via a 
Heisenberg coupling, and this interaction can be controlled by tuning the potential barrier between 
two neighboring dots~\cite{pet05}. Single qubit operations rely on electron spin resonance and 
can be performed by applying local electric or magnetic fields~\cite{kop06,now07}.\\ 
Decoherence -- interaction with the environment -- is mainly 
mediated by two processes, spin-orbit interaction and hyperfine interaction~\cite{han07}. 
Spin-orbit interaction does not have a direct effect on the electron spin, since the electrons do 
not move, but it leads to a mixing of spin and orbital degrees of freedom~\cite{kha01}. In GaAs 
quantum dots, spin-orbit interaction is estimated to be small -- both experimentally~\cite{zum02} and 
theoretically~\cite{gol04} -- compared to the hyperfine interaction with the nuclei, so that the 
latter is the dominant source of dephasing.  

If the atoms of the semiconductor material have a non-zero nuclear magnetic moment (as for example 
in GaAs; in other materials, such as purified SiC, this effect is not present), the electron 
spin $\vec{S}$ interacts with the nuclear spins via the hyperfine interaction~\cite{abr61}: the 
Hamiltonian for such a system can be written as~\cite{kha02, mer02}
\be\label{eHhf}
 \mcH_{hf} = b_{0} S_{z} + \epsilon_{nz} I_{z} + \vec{h} \cdot \vec{S}.
\ee
Here, $b_{0} = g^{*}\mu_{b}B_{0}$ ($\epsilon_{nz} = g_{I} \mu_{n} B_{0}$) is the electron (nuclear) 
Zeeman splitting [calculated using the Bohr (nuclear) magneton $\mu_{B}$ ($\mu_{N}$, where 
$\mu_{N} \ll \mu_{B}$) and the effective $g$-factor of the electron (nuclei), $g^{*}$ ($g_{I}$), 
which in GaAs takes the value $g^{*} = -0.44$]. Next, $I_{z} = \sum_{k=0}^{n-1} \vec{I}_{z}^{k}$ is 
the sum over the $z$-component of all nuclear spins $\vec{I}^{k}$, and $\vec{h} 
= \sum_{k=0}^{n-1}A_{k}\vec{I}_{k}$ denotes the quantum field of the nuclei acting on the electron 
spin, where $n$ are the number of nuclei whose wave function overlaps with the electron's wave 
function ($n \approx 10^{5}$ for typical dots), $A_{k}$ denotes the coupling strength between the 
$k$-th nucleus and the electron. Since the electron's wave function is zero outside the dot, there 
is no overlap with the nuclei outside the quantum dot -- thus each electron in the array couples to 
a different bath of nuclei, and the decoherence is thus local, as stated above. Since hyperfine 
interaction is the dominant source of noise, we can disregard other types of noise which might 
induce some correlations between different qubits, as for example phonons. \\
For an intuitive semiclassical description of decoherence due to hyperfine interaction the quantum 
field of the nuclear spins can be treated as an additional (classical) magnetic field -- the 
Overhauser field -- by replacing $g^{*}\mu_{B}\vec{B}_{n} \equiv \vec{h}$. The maximum value this 
field can reach in GaAs is about~\cite{pag77} $B_{max}=5$~T for fully polarized nuclei. In low 
external magnetic fields, the Overhauser field undergoes Gaussian fluctuations around a 
root-mean-square value~\cite{kha02, mer02, bra05} of $B_{max}/\sqrt{n}$. The electron thus feels 
a total magnetic field which consists of the sum of the controlled external field $\vec{B}_{0}$ and 
the random Overhauser field $\vec{B}_{n}$. The field's longitudinal component $B_{nuc}^{z}$ 
(parallel to $\vec{B}_{0}$) changes the precession frequency of the electron spin by $h_{z} = 
g^{*}\mu_{B}B_{nuc}^{z}$. The transverse part $B_{nuc}^{x,y}$ changes the precession frequency even 
only in quadratic order, $\approx g^{*}\mu_{B} (B^{x,z})^{2}_{nuc}/B_{0}$. This random nuclear 
magnetic field changes in time: two nuclei with different coupling strength $A_{k}$ can exchange 
their spin, thus leading to a change in the Overhauser field $B_{n}$; these fluctuations appear on 
a time scale of $10-100~\mu$s (for a weak external field)~\cite{shu58}, but could probably be 
extended up to well more than several seconds to minutes (for the longitudinal nuclear field 
$B_{nuc}^{z}$ in a strong external field $B_{0}$)~\cite{hut05}. \\
The bulk dephasing time $T_{2}^{*}$ (at which the fluctuating nuclear magnetic field removes the 
phase information) can be measured by rotating the spin in the $xy$-plane, let it evolve freely, and 
then rotate it back along the $z$-axis for measurement (so-called spin echo measurements). Each 
data point then has to be averaged over many measurements, during which the nuclear field evolves. 
This leads to an average dephasing time $T_{2}^{*}$, which has been measured to be about 
$100$~ns~\cite{kik98}.\\
The dephasing time $T_{2}$ of a single electron, on the other hand, is very hard to measure, because 
it is not possible to measure the initial orientation and strength of the nuclear field with 
sufficient precision. Estimates in various regimes predict $T_2 \sim 1 - 100~\mu$s: a good way to 
estimate $T_{2}$ is by using a Hahn echo technique, where the free evolution of the spin due to the 
initial magnetic field is undone by reversing the spin, but not the dephasing due to the change of 
this nuclear field~\cite{her56}. Assuming Gaussian fluctuations of the nuclear field on a time 
scale of $10$~s and (a conservatively estimated) $T_{2}^{*}=10$~ns, a time $T_{2}=10~\mu$s can be 
extracted~\cite{han07}, which has been confirmed by measurements~\cite{pet05} providing a lower 
bound on $T_2$ of $1.2~\mu$s.

For a microscopic, quantum-mechanical treatment of $\vec{h}$, we rewrite $\mcH_{hf}$ 
in~\eqref{eHhf} in a parallel and transversal part~\cite{kha02, mer02, coi04, chi08}
\be\label{eHhfV}
 \mcH_{hf} = \underbrace{(b_{0} + h_{z})S_{z}}_{\mcH_{hf}^{0}} + \underbrace{\frac{1}{2}\left(h_{+}S_{-} + h_{-}S_{+}\right)}_{V}.
\ee
$V$ describes a flip-flop interaction between the electron and a nucleus, thus the operators are 
the raising and lowering operators for the spin ($S_{\pm} = S_{x}\pm i S_{y}$) and a nucleus 
($h_{\pm} = h_{x}\pm i h_{y}$). This perturbation $V$ is small as soon as there is some external 
magnetic field and the energy mismatch between the electron and the nuclear spin states suppresses 
it, as discussed above: expressed in numbers, this requires $|A| \ll 2 |g^* \mu_B \vec{B}_0 - 
g_I \mu_N \vec{B}_{0} + 2 p I A|$~\cite{coi04}, or equivalently (using $\mu_B \gg \mu_N$ and low 
polarisation) $|A| \ll 2 |g^* \mu_B B_0|$, which is fulfilled in typical experiments (with an 
external field above $\gtrsim 3$~T; in Refs.~\cite{elz04,han05}, for example, fields of up to $8$~T 
are used). A first approximation is to completely neglect this term and only consider the change 
of precession frequency due to the nuclear field $h_{z}$. Using the central limit theorem for a 
large number of nuclear spins results in a Gaussian distribution for $h_{z}$. The transverse 
correlator, defined as the self correlation function of the transverse spin component, 
$\mean{S_{+}}_{t} = \tr{e^{i\mcH^0_{hf} t} S_{+} e^{-i \mcH^0_{hf}t} \rho(0)}$ (here, $\rho(0)$ is 
the initial density matrix of the combined system of electron and nuclei), is given 
by~\cite{coi04, chi08}
\be\label{eDecSE}
 \mean{S_{+}^{\textrm{(se)}}}_{t} = \mean{S_{+}}_{0} \exp{-\frac{t^2}{2\tau_{\textrm{se}}^2} + 
 \frac{i}{\hbar} (b+\mean{h_{z}})t}.
\ee
As opposed to exponential decay of phase coherence with time scale $T_2$,~(\ref{eDecSE}) represents 
superexponential 
decay with a characteristic time $\tau_{\textrm{se}} \equiv 2\hbar/A\sqrt{n/(1-p^2)}$: for a GaAs dot 
with almost no polarization, $p\ll 1$, one can estimate $\tau_{\textrm{se}} \approx 5\,\textrm{ns}$,
which is much faster than 
the experimentally observed $T_{2}$-time. The second, imaginary 
part represents the coherent rotation induced by the total magnetic field. The value for 
$\mean{h_{z}}$ depends on the initial state of the nuclei: for a pure state with each 
nucleus having probability $(1+p)/2$ for being in the excited state, it can directly be calculated as 
$\mean{h_{z}} = p A/2$, where $A$ is the hyperfine coupling field.

A more sophisticated approach is to include the perturbation term $V$ in~\eqref{eHhfV}, and 
rewrite the von Neumann equation in the form of a Nakajima-Zwanzig generalized master equation 
(GME)~\cite{coi04}:
\be\label{eNakZwan}
 P\dot\rho(t) = -iPLP\rho(t) - i \int_{0}^{t} \Sigma(t - t')\rho(t') d t'.
\ee
Here $P$ is the projector on the electron-subspace, $L$ the Liouville-operator ($L\mathcal{O} \equiv [\mcH,\mathcal{O}]$ 
for any operator $\mathcal{O}$) and $\Sigma(t)$ the self-energy superoperator. Using regular 
perturbation theory in the parameter $1/b_{0}$ ({\it i.e.} for a high magnetic field $b_{0}\gg A$), 
some (unphysical) secular terms arise; these terms do not occur by directly expanding $\Sigma(t)$ 
in the GME. The latter results in a self correlation function for the transversal spin of the 
form~\cite{coi04} (in the frame oscillating with a frequency proportional to the Zeeman splitting) 
\be\label{eDecBM}
 \mean{S_{+}^{(bm)}}_{t} = \mean{S_{+}}_{t} + R_{+}(t) dt,
\ee
where $\mean{S_{+}}_{t}$ is the Markovian solution, and $R_{+}(t)$ is the remainder term, {\it i.e.} 
the difference between the exponential and the non-markovian solution in Born approximation. This 
can be written as $R_{+}(t) = i \int_{0}^{t} \Psi(t-t')\mean{S^{(bm)}_{+}}_{t'} dt'$, and solved by 
iterating to leading order in the parameter $\delta=A/(4N[b_{0}+h_{z}])$ (corresponding to a high 
external field, since $\delta \sim A/b_{0}$). The solution depends strongly on the wave function of 
the electron: we assume the electron to have a Gaussian wave function in two dimensions, resulting 
in 
\begin{widetext}
\be
\label{eRpl}
 R_{+}(t) \simeq -\delta\mean{S_{+}}_{0}\exp{\frac{itAN}{2\hbar\tau_{\textrm{bm}}(b_{0}+h_{z})}} 
   + \frac{\delta \tau_{\textrm{bm}}^2}{t^2}  \left(  -1+\cos\left[ {\frac{t}{\tau_{\textrm{bm}}}}\right] 
   + p \frac{t}{\tau_{\textrm{bm}}} \sin\left[ {\frac{t}{\tau_{\textrm{bm}}}}\right] 
  + i p\left\{\frac{t}{\tau_{\textrm{bm}}} \cos\left[ {\frac{t}{\tau_{\textrm{bm}}}}\right] 
    -\sin\left[ {\frac{t}{\tau_{\textrm{bm}}}}\right] \right\}  \right).
\ee
\end{widetext}
Here we have defined a characteristic time $\tau_{\textrm{bm}} = 2n\hbar/A$ ($\tau_{\textrm{bm}} 
\approx 1 \mu$s for GaAs quantum dots). In a realistic setting, this correction term 
$R_{+}(t)$ is very small, since $\delta$ is very small: in GaAs typically 
$\delta \approx 10^{-6}$.
Nonetheless, we will calculate this correction for completeness.

Relaxation of an electron spin qubit is caused by the same two effects as dephasing: spin-orbit 
and hyperfine interaction. The required energy for the spin-orbit interaction to flip the spin of 
the electron is provided by the phonons in the lattice of the semiconductors forming the 2DEG, and 
can be calculated as a function of the external magnetic field $B_{0}$~\cite{kha01}. The hyperfine 
contribution to relaxation manifests itself as flips of the electron spin through exchanging its 
spin state with a nuclear spin. For increasing external field, the energy mismatch between the 
nuclear spin states and the electron spin state grows, and more and more energy has to be absorbed 
by phonons -- thus the relaxation can be suppressed by applying a higher external field. The 
relaxation time $T_1$has been measured in experiments to range from $170$~ms (at $B_{0}=1.75\, T$) 
to $120~\mu$s ($B_{0}=14\, T$)~\cite{han03,elz04}.

In a phenomenological model of decoherence, the time scales $T_1$ and $T_2$ are incorporated into 
a master equation model 
for the density matrix with $T_{1}$ on the diagonal (describing the effect of relaxation) and 
$T_{2}$ on the off-diagonal (describing dephasing):
\be
\label{eRhoEv1} 
 \frac{d\varrho}{dt} = \left[\begin{array}{cc}
    (1/T_{1})\, \varrho_{22}  &  -(1/T_{2})\, \varrho_{12} \\
   -(1/T_{2})\, \varrho_{21}  &  -(1/T_{1})\, \varrho_{22} 
 \end{array}\right] .
\ee
Qualitatively, the off-diagonal phase components decrease exponentially with a rate $1/T_{2}$, and 
the ground state $\varrho_{11}$ becomes populated at the expense of the excited state $\varrho_{22}$, 
where  the normalization condition ($\tr{\rho(t)} =~1$) has to be fulfilled at any time $t$. 
(\ref{eRhoEv1}) is a general phenomenological model to describe decoherence, and can thus be 
adjusted to describe decoherence for a wide range of systems, but it does not include microscopic 
information about the quantum processes causing the decoherence.

We now discuss how to extend these decoherence models [Eqs.~\eqref{eDecSE},~\eqref{eDecBM} 
and~\eqref{eRhoEv1}] to more than one qubit. For the exponential decay, this is quite 
straightforward: we rewrite~(\ref{eRhoEv1}) in the Lindblad formalism~\cite{lin76} using the 
Lindblad operator \mcL :
\bea
\label{eRhoLind}
 \mcL\vr & = & \frac{\Gamma_{1}}{2} \left(2\sigma_{+}\vr\sigma_{-} - \sigma_{-}\sigma_{+}\vr - \vr\sigma_{-}\sigma_{+}\right) \nonumber \\
   & & + \frac{\Gamma_{2}}{2} \left(2\sigma_{s}\vr\sigma_{s} - \sigma_{s}\sigma_{s}\vr - \vr\sigma_{s}\sigma_{s}\right).
\eea
Here, $\sigma_{\pm} = 1/2\, (\sigma_{x} \pm i\sigma_{y})$ and $\sigma_{s} = \sigma_{-}\sigma_{+}$ 
are products of the Pauli matrices. Comparing the density matrices resulting from 
Eqs.~\eqref{eRhoEv1} and~\eqref{eRhoLind}, we can identify $\Gamma_{1} = 1/T_{1}$ and $\Gamma_{1} + 
\Gamma_{2}=2/T_{2}$. The time evolution of a single qubit is then found by solving $\mcL\vr(t) 
= d\vr/dt$. To extend~(\ref{eRhoLind}) to multipartite states, we write the Lindblad operator 
for the $k$-th qubit as $\mcL_{k} = \eins\otimes\ldots\otimes\eins\otimes\mcL\otimes\eins\otimes 
\ldots\otimes\eins$, where \mcL\ is the $k$-th operator of a total of $N$
. The time 
evolution of the total $N$-partite state is then given by solving as before $\mcL_N\vr(t) 
= d\vr/dt,$ with $\mcL_{N} = \sum_{k=1}^{N}\mcL_{k}$. By using this definition we implicitly 
assumed that the decoherence of each qubit is governed by the same $\Gamma_{1}$ and $\Gamma_{2}$. 

For the two other models, given in~\eqref{eDecSE} and~\eqref{eDecBM}, we construct the density matrix 
of the entangled states in a similar manner. Since the decoherence of the various qubits is assumed 
to be independent, we can just multiply the corresponding single matrix entries. For statistically 
distributed quantities (as for example $\mean{h_{z}}$), we have to consider the addition rules 
for distributions with the corresponding variances, and furthermore we have to take into account 
which contributions have to be conjugated ({\it e.g.} the precession terms due to the magnetic 
field).
%
%
%
%
\section{Two qubits}
\label{sSit}
Let us first explain our methods and definitions for the simple case of two qubits. A {\it 
separable state} $\vr_{s}$ is defined as a state which can be written as a convex combination of 
product states~\cite{epr35,sch35, werner} 
\be\label{eEntStat}
 \vr_{s} = \sum_{i} p_{i} \vr^{A}_{i}\otimes\vr^{B}_{i},
\ee
where $\vr^{A}$ and $\vr^{B}$ are states in different subsystems $A$ and $B$ and the probabilities 
$p_{i}$ have to fulfill the normalization condition $\sum_{i}p_{i}=1$. If a state is not of this 
form, it is called {\it entangled}.

We use witness operators~\cite{terhal, horo, lew00,guh02,guh03,guh09} to investigate the entanglement of 
various states. An observable $\mcW$\ is called an {\it entanglement witness} if it fulfills the 
following two requirements:
\begin{enumerate}
\item For any separable state $\varrho_{s}$, the expectation value of $\mcW$\ is larger than zero: \\
   $\tr{\mcW \varrho_{s}} \equiv \mean{\mcW}_{\varrho_{s}} \ge 0\ \textrm{for all}\ \varrho_{s}$ separable.
\item There must be at least one entangled state $\varrho_{e}$ for which $\mcW$\ has a negative 
   expectation value:\\ 
   $\textrm{there exists a } \varrho_{e} \ {\rm entangled}\ \textrm{for which}\ \mean{\mcW}_{\varrho_{e}} < 0$.
\end{enumerate}
Therefore, a measured negative expectation value of the witness guarantees that the state is 
entangled. For the experimental implementation, entanglement witnesses can be decomposed into local 
measurements (see also below), and they usually require much fewer measurements than procedures 
such as full state tomography. Thus they are experimentally easier to implement. Finally, it should 
be noted that witnesses can be used to quantify entanglement, by giving lower bounds on 
entanglement measures~\cite{grw}. 

The witnesses we use in this paper are derived from the so-called projector-like 
witness~\cite{bou04}: 
\be \label{eProjWit}
 \mcW_{\psi} = c \eins - \ketbra{\psi} ,
\ee
with the constant $c$ standing for the maximum overlap between the state $| \psi \rangle$ and any 
separable state. Physically, this witness encodes the fact that if a state $\vr$ has a fidelity
$F=\tr{\vr \ketbra{\psi}}$ larger than $c$, then $\vr$ must be entangled.

We first investigate the time evolution of 
the Bell state~\cite{epr35,bell64} $\ket{\Psi^{-}} \equiv \frac{1}{\sqrt{2}}(\ket{01}-\ket{10})$. 
The choice of this Bell-state, the singlet state, is motivated by the fact that it is the ground 
state of the quantum system consisting of two electron spins in a double quantum dot~\cite{han07}, 
thus it is the simplest entangled state that can be created in quantum dots. 

\begin{center}
\begin{figure}[ht]
 \centering
 \scalebox{0.75}{\includegraphics{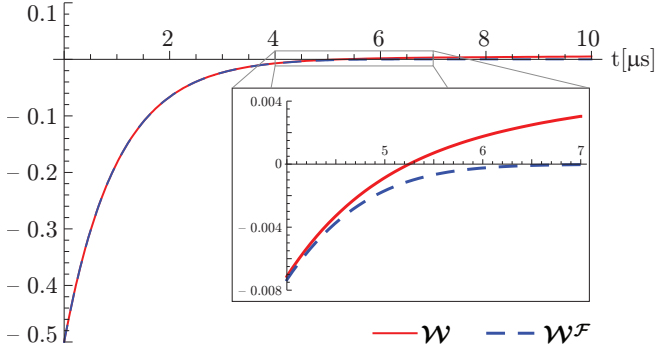}}
 \caption{\label{fBiWit} Expectation values for the regular [~\eqref{eSingWit}, solid red line] and the filtered 
  [~\eqref{WSfilter}, dashed blue line] witness, for $\Gamma_{1} = 10^{3} s^{-1}$ and 
  $\Gamma_{2} = 10^{6} s^{-1}$. The inset 
  shows a zoom into the region where the regular witness becomes positive. $y$ is chosen such 
  that the witness expectation value becomes minimized at any given time, thus it is 
  time-dependent.
 }
\end{figure}
\end{center}

The density matrix of a singlet state under exponential decay can be found from 
Eqs.~\eqref{eRhoEv1} and~\eqref{eRhoLind}:
\be \label{eBiTSt}
 \varrho_{\Psi^{-}}(t) = \frac{1}{2}\left[\begin{array}{cccc}
   2 [1 - \alpha(t)] & 0 & 0   & 0 \\
   0   & \alpha(t) & -\beta(t) & 0 \\
   0   & -\beta(t) & \alpha(t) & 0 \\
   0   & 0         & 0         & 0
 \end{array}\right] ,
\ee
with the factors $\alpha(t) = \exp{- t/T_1} \equiv \exp{- \Gamma_1 t} $ for relaxation and 
$\beta(t) = \exp{- 2 t/ T_2} \equiv \exp{- (\Gamma_1+\Gamma_2) t}$ for dephasing. With that, the 
fidelity~\cite{niel00} $F = \tr{\ketbra{\Psi^{-}}\varrho_{\Psi^{-}}(t)}$ 
is given by 
\be\label{eSingWE}
 F(t) = \frac{1}{2}\left[ \alpha (t) + \beta (t)\right],
\ee
and the expectation value of the projective witness for \ket{\Psi^{-}} is then calculated 
using~\eqref{eProjWit},
\be\label{eSingWit}
 \tr{\mcW_{S} \rho_{\Psi^{-}}(t)}\equiv \mean{\mcW_S}_{\Psi^{-}}(t)  = \frac{1}{2}\left[1 - \alpha(t) - \beta(t) \right] ,
\ee
where $\mcW_{S}$ is the witness for the singlet state, $\mcW_{S}\equiv\eins/2 - \ketbra{\Psi^{-}}$. 
Figure~\ref{fBiWit} shows the decay of entanglement for this exponential model of decay of the coherence.

\begin{center}
\begin{figure}[ht]
 \centering
 \scalebox{0.72}{\includegraphics{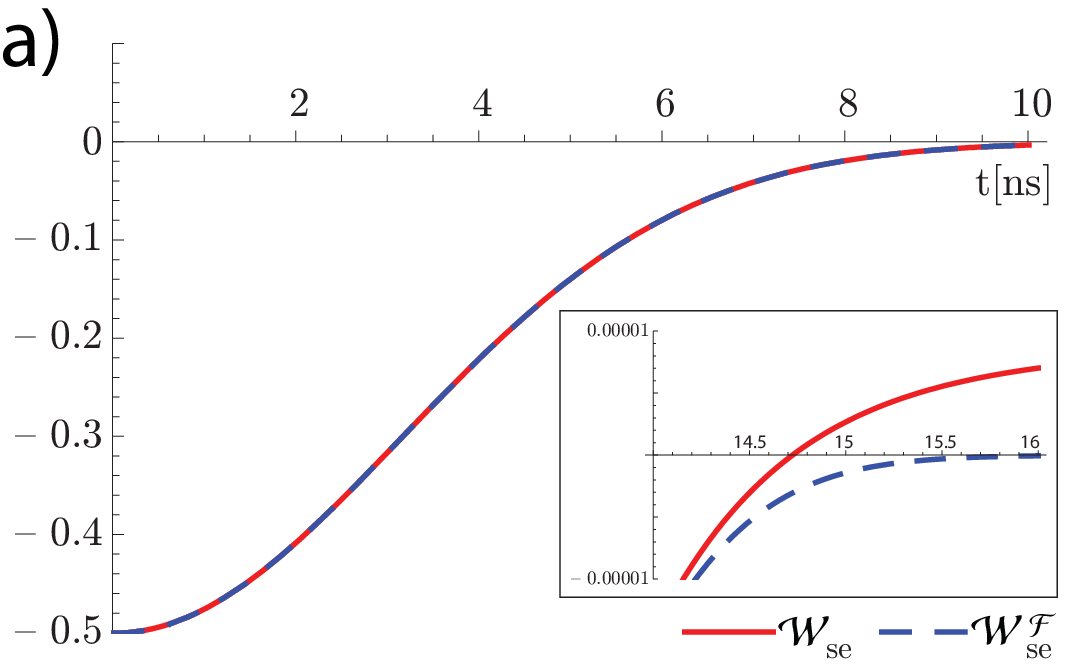}}\hfill
 \scalebox{0.72}{\includegraphics{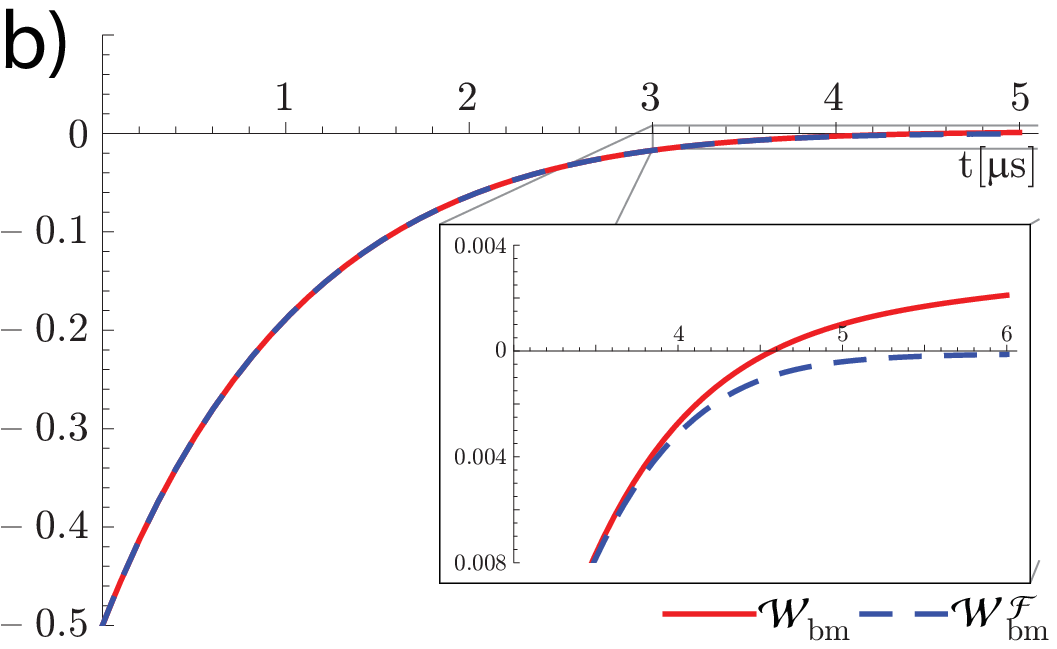}}
 \caption{\label{fWitSpec} a) Decay of the entanglement witnesses [without (solid) and with 
  (dashed) filter] for the superexponential decoherence model using $\tau_{\textrm{se}} = 4.4\,$ns 
  (corresponding to $p = 0.1$). 
  The inset shows the advantage of the filter operator, which works here as well.\\
  b) Decay of the entanglement witnesses for the non-markovian approximation, including the 
  correction term to the Markovian solution, for characteristic time $\tau_{\textrm{bm}} = 1 \mu$s, 
  but an enhanced smallness parameter $\delta = 0.1$ in order to underline the effect of the 
  correction term. 
  }
\end{figure}
\end{center}

For the other two models of decoherence [Eqs.~\eqref{eDecSE} and~\eqref{eDecBM}], we first have to construct the 
corresponding density matrix: for the relaxation, we keep the exponential terms $e^{-t/T_{1}}$, 
but for the dephasing we use the correlators presented in the previous section. The first model is 
based on the superexponential dephasing from~\eqref{eDecSE} for each qubit, whose 
density matrix we label with $\rho_{\textrm{se}}(t)$. We have to calculate the entries $\ket{01}$ 
and $\ket{10}$, therefore the conjugation reverses the phase in~\eqref{eDecSE}, so the 
off-diagonal dephasing terms $[\rho_{se}]_{(2,3)} = [\rho_{se}]_{(3,2)}^{*}$ (the star $^{*}$ 
stands for complex conjugate) are given in terms of the correlators $\mean{S_{+}^{(i)}}_{t}$ of the 
$i$-th dot ($i\in\{1,2\}$) by:
\bea\label{eCorSinSE}
 [\rho_{se}]_{(2,3)}(t) & = & \mean{S_{+}^{(1)}}_{t} \mean{S_{+}^{(1)}}_{t}^{*} \nonumber\\
   & = & \exp{-\frac{\tau_1^2+\tau_2^2}{2\tau_1^2\tau_2^2}t^2+\frac{i}{\hbar}(\mean{h_{z}^{(1)}}-\mean{h_{z}^{(2)}})t} \nonumber\\
   & = & \exp{-\frac{t^2}{\tau_{\textrm{se}}^2}},
\eea
where the second line is for identical statistics of the dots (thus with identical characteristic 
times $\tau_1 = \tau_2 \equiv \tau_{\textrm{se}}$).
Including the boundary condition $[\rho_{se}]_{(2,3)}(0) = -1/2$ we obtain the density matrix: 
\be\label{density_se}
 \rho_{se}(t) = \frac{1}{2}\left[\begin{array}{cccc}
   2 [1 - \alpha(t)] & 0 & 0   & 0 \\
   0   & \alpha(t) & -\beta_{se}(t) & 0 \\
   0   & -\beta_{se}^{*}(t) & \alpha(t) & 0 \\
   0   & 0         & 0         & 0
 \end{array}\right] ,
\ee
with $\alpha(t)=\exp{-t/T_{1}}$ as before and $\beta_{se}(t) = [\rho_{se}]_{(2,3)}(t)$ from~\eqref{eCorSinSE}. 
The witness operator $W_{se}$ for detecting (\ref{density_se}) is thus 
the same as in~\eqref{eSingWit} with the replacement $\beta(t)
\rightarrow\beta_{se}$. The evolution of the 
corresponding witness is shown in figure~\ref{fWitSpec} a).

The third model uses the non-markovian Born approximation for the decay, Eqs.~\eqref{eDecBM} and~\eqref{eRpl}. 
The single electron decay is given by $\mean{S_{+}^{\textrm{(bm)}}}_{t} = 
\exp{-t/(2 T_{1}+T_{2})} + R_{+}(t)$, where the first exponential term stems from the Markovian 
solution, and the remainder term is given in~\eqref{eRpl}. In order to set up the 
density matrix $\rho_{\textrm{bm}}(t)$ in the non-markovian approximation, we replace $\beta(t)$ in~\eqref{eBiTSt} by $\beta_{\textrm{bm}}(t) 
= \mean{S_{+}^{\textrm{(bm)}}}_{t} \mean{S_{+}^{\textrm{(bm)}}}_{t}^{*}$, in the same way as in the 
super-exponential case.

In the next part we introduce a systematic method to enlarge the time interval during which entanglement 
can be detected by the witness operators. This method is based on applying 
local (so-called filtering) operators to the witness operator and analyzing the measurement results in a 
different way without requiring more measurements. Analyzing the witness~\eqref{eSingWit}, we see 
immediately that it becomes positive (and hence does not detect the entanglement anymore) when 
after some finite time $\beta(t)$ becomes smaller than $1-\alpha(t)$, though it can be shown (by 
virtue of the PPT-criterion~\cite{per96}, for example) that the state $\varrho_{\Psi^{-}}(t)$ is 
entangled for any $t < \infty$, {\it i.e.} for any $\beta(t)>0$. 

Therefore, our goal is to construct a witness operator which is able to detect the entanglement in 
the state $\varrho_{\Psi^{-}}(t)$ at any time. This can be achieved by a filter operator $\mcF$,
\be
\label{eFilDef}
\mcF = \mcF_{1} \otimes \mcF_{2},
\ee
where the $\mcF_{i}$ are arbitrary invertible matrices acting on {\it individual} qubits. Since 
\mcF\ is local, application of such a filter operator on a state $\rho$ does not change its 
entanglement properties, \emph{i.e.} $\mcF^{\dagger} \vr \mcF$ is entangled, iff $\vr$ is entangled. 

Equivalently, one can apply filter operators to witness operators $\mcW$, and 
the resulting \emph{filtered witness operator} $\mcW^{\mcF}$ is then given by
\be
 \mcW^{\mcF} = \mcF \mcW \mcF^{\dagger} .
\ee
As normalization, we choose $\tr{\mcW} = \tr{\mcW^{\mcF}}$, to make the witnesses' mean values 
comparable.

Our goal is now to design a filter $\mcF_{i}$ such that it increases the negativity of the 
witness, \emph{i.e.} it should increase the weight of the terms $\alpha$ and $\beta$ 
in~\eqref{eSingWit}, so that the filtered witness can be used to detect entanglement during longer 
times. This can be achieved by the following filter:
\be\label{eFilMat}
 \mcF_{i} = \left[\begin{array}{cc}
    1 & 0 \\
    0 & y
  \end{array}\right],
\ee
with $y$ a positive real number. The normalized filtered witness for the singlet state then takes the form 
\bea \label{eFilWitSing}
  \mcW^{\mcF}_{S} &=& \left(\mcF_{1}\otimes\mcF_{2}\right) \mcW_{S} \left(\mcF_{1}\otimes\mcF_{2}\right)^{\dagger}  
\nonumber
\\
&
=& \frac{1}{1+y^{4}}\left[\begin{array}{cccc}
    1 & 0     & 0     & 0 \\
    0 & 0     & y^{2} & 0 \\
    0 & y^{2} & 0     & 0 \\
    0 & 0     & 0     & y^{4}
  \end{array}\right],
\eea
and the expectation value is given by 
\be\label{WSfilter}
 \mean{\mcW^{\mcF}_{S}}_{\Psi^{-}(t)} = \frac{1}{1+y^{4}}\left[ 1 - \alpha(t) - y^{2}\beta(t)\right] .
\ee
Clearly, $\mean{\mcW^{\mcF}_{S}}_{\Psi^{-}(t)}$ is negative if $y$ is chosen large enough and 
time dependent and $t<\infty$ (thus $\beta (t) > 0$), and the negativity of the witness can be 
optimized by a suitable choice of $y \equiv y(t)$ for a given time $t$. The remaining entanglement 
[which is not detected by $\mcW_{S}$,~(\ref{eSingWit})] in the decohering state can then be 
detected by measuring this filtered witness operator. The effectiveness of the filter operator 
crucially depends on the choice of the singlet state\ket{\Psi^{-}} as the initial state: it can 
easily be shown that for the other Bell states \ket{\Phi^{\pm}} = $\frac{1}{\sqrt{2}}(\ket{00} \pm 
\ket{11})$, the filtered witness does not lead to any improvement over the regular witness 
operator. The decay of entanglement in our model thus strongly depends on the initial state, even 
within the same basis.  

For the experimental implementation, the witness $\mcW^{\mcF}_{S}$ can be decomposed into single-qubit 
measurements~\cite{guh03}:
\begin{align}
&\mcW^{\mcF}_{S} = \frac{1}{4(1+y^4)}[(1+y^{4})(\unity\otimes\unity+\sigma_{z}\otimes\sigma_{z}) 
\\
\nonumber
&-(1-y^{4})(\sigma_{z}\otimes\unity+\unity\otimes\sigma_{z}) + 2y^{2}(\sigma_{x}\otimes\sigma_{x}+\sigma_{y}\otimes\sigma_{y})] .
\end{align}
This decomposition requires three measurement settings (namely $\sigma_{i}\otimes\sigma_{i}$ with 
$i\in\{x,y,z\}$) instead of the nine settings full state tomography would require~\cite{guh02,optDec}. 
Similar decompositions exist for all other witnesses occurring in this paper~\cite{bou04,ijtp}.

In Figs.~\ref{fBiWit} and~\ref{fWitSpec}, the evolution of the expectation values for both the 
regular witnesses (solid line) and the filtered witnesses (dashed line) are plotted. In the 
experimentally relevant limit ($\Gamma_{2} \gg \Gamma_{1}$), the advantage of the filter operator 
in an experiment does not manifest itself as strongly as would be the case for $\Gamma_{2} \simeq 
\Gamma_{1}$; however, the principle advantage that the entanglement can be detected for any finite 
time is demonstrated in the insets by a zoom into the region where the unfiltered witness becomes 
positive. The filtered witness remains negative, albeit with a small, exponentially decaying 
absolute value, for any finite time: this proves that the \ket{\Psi^-}\ contains at least a very 
small amount of entanglement at any time under our decoherence models; however, it will not lead to 
a significant advantage in an experiment, since the noise due to imperfect state preparation and 
measurement fidelities will render it virtually impossible to measure the expectation value with 
such a high precision.

From this curve, one can conclude that it becomes difficult to detect entanglement after more than 
a few $\mu$s under realistic conditions assuming any of the decay models we have considered (for 
the superexponential decay even after a few ns, but for all models the exact time also depends on 
the size of the error bars in a given experiment). Consequently, any generation scheme for the Bell 
states which requires a generation time longer than this time will probably not work in practice.

\begin{center}
\begin{figure}[ht]
 \centering
 \scalebox{0.78}{\includegraphics{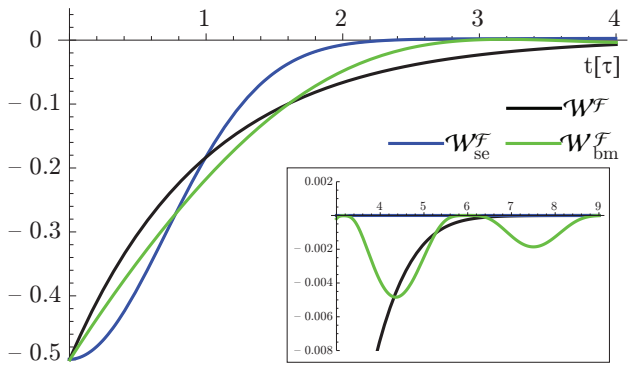}}
 \caption{\label{fWitComp} Comparison of the evolution of the filtered witness operators for 
  exponential decay, superexponential decay, and the Born approximation, using $\delta = 0.1$. 
  The unit of time $\tau$ is the critical time in 
  each model (thus $\tau = T_{2},\, \tau_{\textrm{se}},\, \tau_{\textrm{bm}}$). \\ 
  For short times, both the superexponential and the non-markovian approximation are decaying less 
  strongly than the exponential model, but after some time, the witness assuming exponential decay 
  has a greater negativity. The inset compares the longer-time behavior of the three models: the 
  non-markovian approximation features a periodic recurrence of negative values (due to precession around the 
  nuclear magnetic field), and the slowest long-time decay (disregarding the coherent precession).
 }
\end{figure}
\end{center}

Let us conclude this section by a comparison of the three models: The evolution of the expectation 
values of the filtered witnesses is plotted in figure~\ref{fWitComp} (all in units of the respective 
critical time for the comparison). In the non-markovian approximation, we have chosen the smallness parameter 
$\delta = 0.1$ such that the effect of the correction can be shown. This comparison reveals 
benefits and drawbacks of each decay model: first, the filtered entanglement operators are all 
negative for any finite time $t < \infty$. The entanglement in the superexponential model decays 
faster than in the other two, and in the non-markovian approximation, we see the effect of the power-law tail 
[the second term in~\eqref{eRpl}] as a periodic rebouncing due to the precession around the 
$z$-component of the nuclear magnetic field, $h_{z}$. As a result, the expectation value of this 
witness operator is more negative than for the purely exponential decay. In the main plot showing 
the short time evolution scaled by the critical time for each model, the differences between the 
models are not as pronounced, and they behave roughly the same. \\
Let us consider a realistic experiment at this point: In an experiment, it is likely that errors 
will arise due to imperfect read-out of the electron spin states~\cite{elz04,han05}, which will 
manifest themselves as error bars on the curves for the time evolution. This error will make it 
unlikely to detect the entanglement at longer times in a realistic experiment. Regarding 
figure~\ref{fWitComp} with these errors in mind, the evolution within the three models is 
roughly equivalent. Another source of possible experimental inaccuracies is the preparation of the 
initial state -- in general it will not be the exact aimed-for state, but a mixture of states. This 
mixture will influence the use of the witness and the filter operator, which as well strongly 
depends on the nature of the mixture; though the precise influence is hard to predict, the filter 
will always improve the witness operator to some degree. However, the creation of pure singlet 
states in double quantum dots has already been experimentally achieved in a controllable manner and 
with high probability of success~\cite{pet05}. Therefore we expect that our noiseless results can 
nevertheless be used to give qualitative predictions of the decay of entanglement. \\
So far, we have considered two entangled qubits and found that their entanglement remains 
detectable for about the same time for three decoherence models. When considering generalization to 
many qubits, we note that the exponential model features a big advantage compared to the other two, 
since for this model there is a general method for calculating the time evolution of the density 
matrix for an arbitrary number of qubits (see also Sec.~\ref{sNQub}), whereas for the two other 
models we have to construct the density matrix for each new state by hand. Therefore, in the 
following sections, we will use the exponential model for the generalization to multiple qubits.
%
%
%
%
\section{Three Qubits}
\label{s3Qub}

For three or more particles, the situation is
more complicated, since different classes of 
multiparticle entanglement exist~\cite{ghz89,dur00,aci01}.

Let us first discuss the notion of partial separability. 
A state can be {\it partially separable}, meaning that some of the 
qubit states are separable, but not all. An example for three particles 
is the state 
\be
\ket{\psi^{bs}} = \ket{\phi^{AB}}\otimes\ket{\phi^{C}},
\ee 
where $\ket{\phi^{AB}}$ is a (possibly entangled) state of two qubits 
(defined on subsystems $A$ and 
$B$), and $\ket{\phi^{C}}$ a state of the third qubit (defined on subsystem $C$). 
The state $\ket{\psi^{bs}}$ 
is separable with respect to a certain bipartite split, so it 
is called biseparable. A mixed state is biseparable, if it can be written as mixture
of biseparable pure states.

If a state is not biseparable, it is genuinely multipartite entangled.
There exist different classes of multipartite entangled states~\cite{dur00} 
and the number of entanglement classes increases with the number of qubits~\cite{ver02}. 
An entanglement class can be defined by the following question: given a single copy of 
two pure states 
$\ket{\psi}$ and $\ket{\phi}$, is it possible, at least in principle, to transform
$\ket{\psi}$ into $\ket{\phi}$ (and vice versa) using local transformations only?
Even if the probability of success is small?
For three qubits, for example, two entanglement classes exist, the GHZ and the W-class. 
Every genuine multipartite entangled three-qubit state can be transformed into 
one of the two states~\cite{dur00} 
\bea
 \ket{GHZ_{3}} & = & \frac{1}{\sqrt{2}} \left( \ket{010} + \ket{101} \right) , \label{eGHZ3}\\
 \ket{W_{3}}   & = & \frac{1}{\sqrt{3}} \left( \ket{100} + \ket{010} + \ket{001} \right) \label{eW3},
\eea
but, remarkably,  these two states cannot (not even stochastically) be transformed into 
each other, and are therefore representatives of different entanglement classes. 

Let us now investigate the lifetime of these two states using our exponential decoherence model,
described below~(\ref{eRhoLind}). After calculating the time evolution of the two states, 
we obtain for the corresponding fidelities:
\bea
 F_{GHZ}(t) & = & \frac{1}{4} \Big(\exp{-2\Gamma_{1}t} + \exp{-\Gamma_{1}t} 
\nonumber
\\
&& +2\,\exp{-\frac{3}{2}(\Gamma_{1}+\Gamma_{2})t}\Big) ,
\label{eFidGHZ3}\\
 F_{W}  (t) & = & 
\frac{1}{3} \left(\exp{- \Gamma_{1}t} + 2\, \exp{-(\Gamma_{1}+\Gamma_{2})t} \right). 
\label{eFidW3} 
\eea
{From} these fidelities, the expectation values of the witnesses can directly be determined
as $\mean{\mcW_{G}}_{\rho_{G}(t)} \equiv 1/2 -F_{GHZ}(t) $ for the GHZ state, and 
$\mean{\mcW_{W}}_{\rho_{W}(t)} \equiv 2/3 -F_{W}(t)$ for the W state. 

\begin{center}
\begin{figure}[ht]
 \centering
 \scalebox{0.74}{\includegraphics{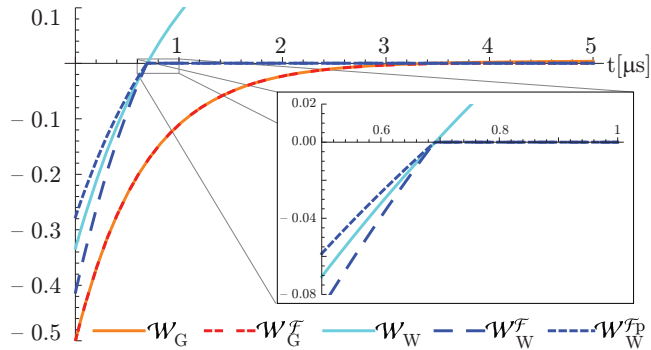}}
 \caption{\label{f3Wit} Expectation values for the witnesses for the tripartite GHZ- and W-states, 
  both regular (solid lines) and filtered (dashed), as a function of time $t$. The projected 
  filter $\mcW_{W}^{\mcF p}$ (see text) is plotted as well, though it is not better than the filtered 
  witness $\mcW_{W}^{\mcF}$. Parameters used are $\Gamma_{1} = 10^{3} s^{-1}$ and $\Gamma_{2} = 
  10^{6} s^{-1}$. As in figure~\ref{fBiWit}, $y$ is time dependent and chosen so as to minimize 
  the witness expectation value.
 }
\end{figure}
\end{center}

Our next step is to apply the filter operators to the witnesses. This yields the following 
values for the expectation value of the filtered witness operators: 
\begin{widetext}
\bea
 \mean{\mcW_{G}^{\mcF}}_{\varrho_{G}(t)} & = & \frac{3}{2 \left(1+2 y^2+2 y^4+y^6\right)} \left[ 2 +  \left(-3+2 y^2\right) e^{-\Gamma_{1}t}
      + \left(1-2 y^2\right) e^{-2\Gamma_{1}t} - 2 y^3 e^{-\frac{3}{2}(\Gamma_{1}+\Gamma_{2})t} \right] , \label{eG3Wit} \\
 \mean{\mcW_{W}^{\mcF}}_{\varrho_{W}(t)} & = & \frac{13}{3 \left(2+3 y^2+6 y^4+2 y^6\right)} 
 			\left[2 + (-2+y^2) e^{-\Gamma_{1}t}
 			- 2 y^2 e^{-(\Gamma_{1}+\Gamma_{2})t} \right] . \label{eW3Wit}
\eea
\end{widetext}
These witness operators can be measured by four (for the GHZ state) or five (for the W state) 
measurement settings~\cite{ijtp}, compared to the 27 measurement settings required for full state 
tomography. In principle, the witness for the W state can be improved by taking the projector onto 
the subspace with at most two excitations~\cite{guh09}, ($\eins_{2} = \eins - \ketbra{111}$ instead 
of $\eins$). However, in the present case this does not give any improvement, since the 
$\ketbra{111}$ state is not populated. The time evolution of the witness expectation values 
(\ref{eG3Wit}) and (\ref{eW3Wit}) are plotted in figure~\ref{f3Wit} (for $\Gamma_{1} = 
10^3$~s$^{-1}$ and $\Gamma_{2} = 10^6$~s$^{-1}$).

\section{Four Qubits}
\label{s4Qub}
The more qubits are added, the more distinct classes of entangled states arise. 
For four qubits, we investigate the following four classes:
\begin{align}
 \ket{GHZ_{4}}  = &  \frac{1}{\sqrt{2}} \left(\ket{0101} + \ket{1010} \right) \label{eGHZ4} , \\
 \ket{C_{4}}    = &  \frac{1}{2}        \left(\ket{0101} + \ket{0110} + \ket{1001} - \ket{1010} \right) \label{eClu4} , \\
 \ket{W_{4}}    = &  \frac{1}{2}        \left(\ket{1000} + \ket{0100} + \ket{0010} + \ket{0001} \right) \label{eW4} , \\
 \ket{D_{4}}    = &  \frac{1}{\sqrt{6}} (\ket{0011} + \ket{0101} + \ket{0110} 
\nonumber
\\
& +
 \ket{1001} + \ket{1010} + \ket{1100} ) \label{eDic4} . 
\end{align}
All of these states have been realized in various experiments for different physical 
systems~\cite{4qbExp}, but so far not in solid-state nanosystems. Also, some of their
decoherence properties have been investigated from different theoretical 
perspectives~\cite{
decoher,guh08}. The states \ket{GHZ_{4}} and \ket{W_{4}} 
are the four-qubit versions of the states we have investigated 
for three qubits in the previous section. \ket{C_{4}} is a representative of the so-called cluster 
class~\cite{rau01}, important in the context of one-way quantum computing~\cite{hei05}. 
The Dicke state~\cite{dic54} \ket{D_{4}} is an extension of the W-state and consists of
all possible permutations of states containing 2 excitations.
The fidelity of these states evolves as:
\begin{widetext}
\bea
 F_{GHZ4} (t) & = & \frac{1}{2} \left( \exp{-2\Gamma_{1}t} + \exp{-2(\Gamma_{1}+\Gamma_{2})t} \right) , \label{eFidGHZ4} \\
 F_{C4}   (t) & = & \frac{1}{2} \left( \exp{-2\Gamma_{1}t} + \exp{-2(\Gamma_{1}+\Gamma_{2})t} + \exp{-(2\Gamma_{1}+\Gamma_{2})t} \right) ,
 \label{eFidC4} \\
 F_{W4}   (t) & = & \frac{1}{4} \left( \exp{- \Gamma_{1}t} + 3\, \exp{-(\Gamma_{1}+\Gamma_{2})t} \right) , \label{eFidW4} \\
 F_{D4}   (t) & = & \frac{1}{6} \left( \exp{-2\Gamma_{1}t} + \exp{-2(\Gamma_{1}+\Gamma_{2})t} + 4\, \exp{-(2\Gamma_{1}+\Gamma_{2})t} \right) .
 \label{eFidD4}
\eea
\end{widetext}
The corresponding projective witnesses can be found, as before, using $c\eins - F(t)$, with $c=1/2$ for the 
cluster and GHZ-states, $c=3/4$ for the W-state, and $c=2/3$ for the Dicke state~\cite{tothdicke}. 

\begin{center}
\begin{figure}[ht]
 \centering
 \scalebox{0.78}{\includegraphics{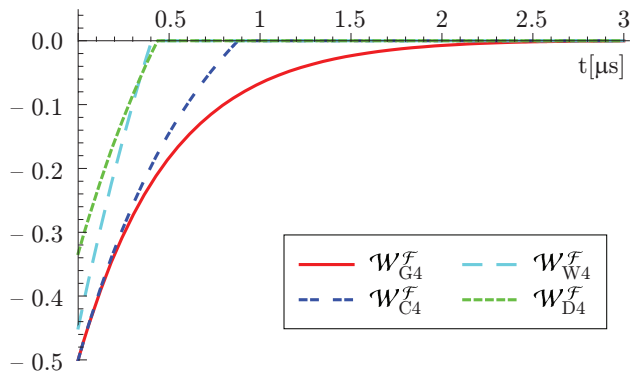}}
 \caption{\label{f4Wit} Expectation values for the filtered witness operators 
  for the four considered classes of fourpartite entangled 
  states (GHZ, cluster, W and Dicke). The filter operator makes the entanglement in the GHZ-state 
  detectable for arbitrarily long times (at least in principle). 
  The entanglement of the other 
  three classes decays faster, but the cluster state is more stable (decays slower) than both the W- 
  and the Dicke states. Parameters used are the same as in figure~\ref{f3Wit}.
 }
\end{figure}
\end{center}

Again, filter operations can be applied: the resulting formulas are lengthy and therefore not given 
here. The improvement over the regular witness again shows (as for the general case of $N$ qubits) 
that the GHZ-state contains in theory entanglement for any finite time -- but so little, that this 
result is of a theoretical nature and not experimentally relevant. For the other classes of states, 
the filter can lead to a slightly higher negativity, but not to an extension of the time where the 
expectation value will become positive. So what is left is to compare the differences in the 
evolution of the expectation values of these witness operators for the four classes, and to see 
which one is the most stable, \emph{i.e.} detectable for the longest time. This is done in 
figure~\ref{f4Wit}. 

At this point the same question can be asked as for the two-qubit state that we 
investigated in Sec.~\ref{sSit}: how does the available detection time depend 
on the exact state chosen as representative of a class? Or, equivalently, the fidelity 
of which state decoheres most slowly? In fact, writing the states above in a different
basis leads to different decay rates.

\begin{center}
\begin{figure}[ht]
 \centering
 \scalebox{0.78}{\includegraphics{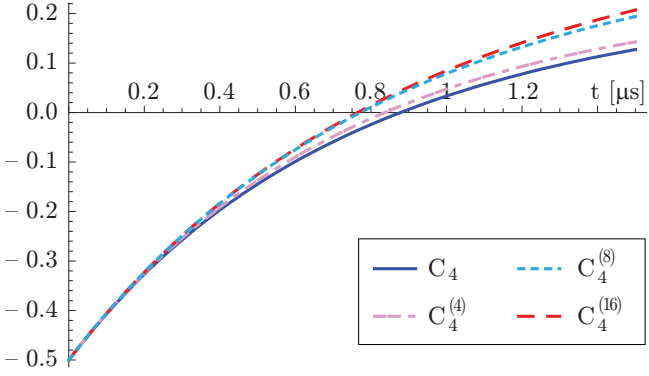}}
 \caption{\label{f4BasDep} Expectation values of the witnesses for some four-partite cluster states. 
The upper index 
  indicates the number of terms in the representation of each state, the state without index is the 
  one given in~\eqref{eClu4}. Parameters used are the same as in figure~\ref{f3Wit}.
 }
\end{figure}
\end{center}

This is illustrated in figure~\ref{f4BasDep}, where the evolution of the witness 
expectation value of four different cluster states is plotted: \ket{C_{4}}\ 
from~\eqref{eClu4}, \ket{C_{4}^{(16)}} is the original cluster state from Ref.~\cite{bri01} 
containing 16 terms, and the two additional representations 
\bea
 \ket{C_{4}^{(4)}} & = & \frac{1}{2} \left( \ket{0000} + \ket{0011} + \ket{1110} + \ket{1101} \right), \label{eClu44} \\
 \ket{C_{4}^{(8)}} & = & \frac{1}{2^{3/2}} \left( \ket{0000} + \ket{0011} + \ket{0100} + \ket{0111} \right. \nonumber\\
 										&   & \left. - \ket{1000} - \ket{1010} + \ket{1101} + \ket{1110} \right) \label{eClu48},
\eea
with 4 and 8 terms, respectively (the first one is a representation with the minimal number of 
terms, which will be used again in the next section, the second one a rotated version of the 
original cluster state). As can be seen in figure~\ref{f4BasDep}, the detection time decreases as the 
number of terms increases, though the effect is not very large for four and more terms. The representations
with the minimal number of terms thus decohere more slowly. This  is not surprising, since one can prove 
for a similar decoherence model that states with the minimal number of terms are most robust~\cite{guh08}.
For representations with the same number of terms, the number of excitations in each term can influence 
the detectability: which one of the two is easier to detect then depends on the ratio of 
$\Gamma_{1}$ and $\Gamma_{2}$.

%
%
%
\section{N qubits}
\label{sNQub}

Let us now consider the general situation of $N$ qubits. We concentrate on three types of entangled states
for which there exist proposals 
how to generate them using available single- and two-qubit operations in quantum 
dots~\cite{bod07}:  
GHZ-, W-, and cluster states. Our goal is to calculate the time evolution 
of the normal (unfiltered) witness for arbitrary $N$ and compare this with the time necessary 
to generate and measure the state. The representatives of the first 
two classes can be written down straightforwardly (for even $N$):
\begin{align}
 & \ket{GHZ_{N}} =  \frac{1}{\sqrt{2}} \left( \ket{01\ldots 01} + \ket{10\ldots 10}\right) , \label{eGHZN}\\
 & \ket{W_{N}}   =  \frac{1}{\sqrt{N}} \left( \ket{00\ldots 01} + \ket{00\ldots 10} + \ldots + \ket{10\ldots 00} \right) . \label{eWN}
\end{align}
Calculating the expectation value of the witness operators leads to (for even $N$): 
\begin{align}
 \mean{\mcW_{G}}_{\varrho_{G}(t)} =
  & \frac{1}{2} \left\{ 1 - \exp{-\frac{N}{2} \Gamma_{1}t} - \exp{-\frac{N}{2} (\Gamma_{1}+\Gamma_{2})t} \right\}, \label{eGNWit} \\
 \mean{\mcW_{W}}_{\varrho_{W}(t)} =
  & \frac{1}{N} \big\{ N-1 - \exp{-\Gamma_{1}t}  \nonumber \\
  & -(N-1)\exp{-(\Gamma_{1}+\Gamma_{2})t} \big\} \label{eWNWit}
\end{align}

The general form of the cluster state -- the one we consider here containing the minimal number of 
terms, namely $2^{N/2}$ -- is more complicated; it can be written as~\cite{clustExp}
\be\label{clusterlang}
 \ket{C_{N}} = \bigotimes_{k=1}^n \frac{[\ket{00}+\ket{11}(\sigma_x \otimes \eins)]}{\sqrt{2}},
\ee
where this formula should be understood as an iteration, with the operator 
$(\sigma_x \otimes \eins)$ acting on the Bell state of the next two qubits. 
For four qubits, this results exactly in the representation \ket{C_{4}^{(4)}} 
from~\eqref{eClu44}, the evolution of which is plotted in figure~\ref{f4BasDep}. 
To calculate the time evolution of the fidelity, we represent the cluster state 
(\ref{clusterlang}) as~\cite{hei05}
\be \label{clusterprod}
\ketbra{C_{N}} = \prod_{k=1}^N \frac{\eins + S_k}{2}
\ee
with $S_k$ a product of Pauli matrices. We incorporate 
the effects of dephasing (disregarding the relaxation of the qubits, i.e. 
setting $\Gamma_{1} = 0$, which leads to 
an error of less than 
$0.01$\textperthousand\ for four qubits)
in every term in the sum of the expanded~(\ref{clusterprod}). The resulting fidelity of the
cluster state can then easily be calculated numerically up to $N = 24$ qubits.

\begin{center}
\begin{figure}[ht]
 \centering
 \scalebox{0.78}{\includegraphics{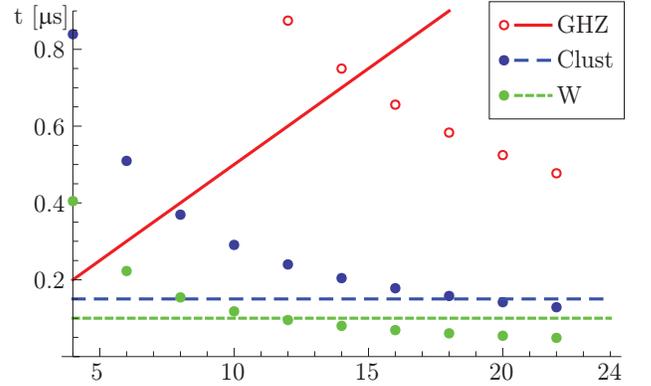}}
 \caption{\label{fNWit} The dots show the time at which the expectation value of the (unfiltered) 
 witness for a given class of states becomes positive 
  as a function of the qubit number $N$. Parameters used are $\Gamma_{1} = 10^{3} s^{-1}$ and  
  $\Gamma_{2} = 10^{6} s^{-1}$, except for the cluster state where the relaxation 
  is disregarded ($\Gamma_{1} = 0$). Also plotted are (shown by lines) the times required to 
  generate the given entangled states, see the text for explanation. 
 }
\end{figure}
\end{center}

Figure~\ref{fNWit} shows the time at which the expectation value of the (unfiltered) projective 
witness for each of the three states [Eqs.~\eqref{eGNWit}, \eqref{eWNWit} and~\eqref{clusterprod}]
becomes positive 
as a function of the qubit number $N$, as well as a rough estimate of the time 
necessary to generate and measure these states. For electron spin qubits in quantum dots 
the generation times are taken from Ref.~\cite{bod07}: for both the cluster and 
the W-states the time required to produce these states
is independent of the number of qubits, whereas the production time of states of the 
GHZ-class scales linearly with the number of qubits. The measurement times are 
composed as follows: measurement 
distinguishes between spin-up and spin-down (defined along the 
z-axis~\cite{elz04}) and measuring the 
components $\sigma_{x}$ and $\sigma_{y}$ then requires a rotation 
of the spins by $\pi/2$, which takes about $\sim 50$ ns~\cite{kop06}. 
The sum of the generation and the measurement time is given by the lines in the plot. 

We see in figure~\ref{fNWit} that (as in Figs.~\ref{f3Wit} and ~\ref{f4Wit}) the entanglement of 
the GHZ-state can be detected for the longest times, but it is more time-consuming to generate 
than the other two entangled states. Based on the estimates in figure~\ref{fNWit}, generation and 
detection of GHZ states should be possible for up to 14 qubits (with the standard projective 
witness and assuming current operation and decoherence times for electron spin qubits). The cluster 
state is the state which can be detected for the largest number of qubits, although for up to 
12~qubits the ``time reserve'' (\emph{i.e.} the difference between the time needed for generation and 
measurement and the time when the expectation value of the witness operator becomes positive) for 
the GHZ state is somewhat larger than for the cluster state. The W-state is the least suitable, 
the largest state would contain about $\sim 10~$~qubits. 

Our results for the cluster state show 
that one-way quantum computing~\cite{rau01} is not really feasible in quantum dots with current 
dephasing times: we expect that up to maximally $\sim 12$~qubits could be entangled under the 
presented preparation scheme, which is far too few for exploiting the advantages of a quantum 
computer.

Based on our assumptions, thus, the simplest state to generate and prove it's entanglement would 
be the GHZ-state for up to twelve qubits, and the cluster state for more than twelve and up to 
twenty qubits, though the remaining entanglement becomes very small. The same holds for the 
filtered witness for the GHZ state for an arbitrary number of qubits.
\section{Conclusion}\label{sConc}
In conclusion, we have investigated entanglement and its detectability in a linear array of 
electron spin qubits 
which locally undergo decoherence. We have considered three different phenomenological models for the 
dephasing of the qubits based on exponential and superexponential decay.
Using witness operators as detectors of entanglement and introducing a specific class of 
filtered witness operators, we estimated the maximum available detection time for entanglement
of two electrons using each of the models and found that the time during which entanglement is 
detectable is independent of the model chosen. We then expanded the exponential model to the case 
of multipartite entanglement: For three and four qubits, we compared the decay of entanglement for 
different classes of entangled states with each other, namely the GHZ-, W-, cluster and Dicke 
classes. We also gave limits on the maximum number of entangled qubits that can be created and 
measured based on currently known decoherence times for electron spin qubits. The most suitable 
entangled state turns out to be the GHZ-state for up to a few qubits. Our results can help to make 
a choice as to which state to prepare in experiments. Since local decoherence is characteristic for 
many types of solid-state qubits, our model and the filtered operator technique are applicable to a 
variety of these qubits.

This work has been supported by The Netherlands Organisation 
for Scientific Research (NWO), the FWF (START prize) and the 
EU (SCALA, OLAQUI, QICS).


\begin{thebibliography}{99}
\bibitem{clustExp}
In general, a cluster state is defined as an eigenstate of certain local observables~\cite{rau01,hei05}; 
here we choose them as 
$\mcS_{1}=\sigma_{z}\sigma_{z}\eins\ldots$; $\mcS_{2}=\sigma_{x}\sigma_{x}\sigma_{x}\eins\ldots$; 
$\mcS_{3}=\eins\sigma_{z}\sigma_{z}\sigma_{z}\eins\ldots$; $\ldots$; $\mcS_{N}=\eins\ldots\eins\sigma_{x}\sigma_{x}$. 
This is locally equivalent to the definition as in~\cite{rau01,hei05}; however, the state 
given in~\eqref{clusterlang} contains the minimal number of terms; see Ref.~\cite{guh08}.
\end{thebibliography}
\end{document}